\documentclass[cameraready]{Interspeech}

\title{G-MaP-SE: Guided Speech Enhancement via GMM-Based Prior Matching}

\author[affiliation={1}]{Yike}{Zhu}
\author[affiliation={1}]{Ziqian}{Wang}
\author[affiliation={1}]{Zikai}{Liu}
\author[affiliation={1}]{Xingchen}{Li}
\author{Zhuangqi}{Chen}
\author{Xianjun}{Xia}
\author{Chuanzeng}{Huang}
\author[affiliation={1}, correspondingauthor]{Lei}{Xie}

\address{
    $^1$ Audio, Speech and Language Processing Group (ASLP@NPU),\\
    School of Software, Northwestern Polytechnical University, Xi'an, China
}

\email{ykzhu@mail.nwpu.edu.cn, lxie@nwpu.edu.cn}


\usepackage{comment}
\usepackage{amsmath}
\usepackage{mathtools}
\let\|\relax
\DeclarePairedDelimiter{\norm}{\lVert}{\rVert}
\usepackage{graphicx}
\usepackage{hyperref}
\usepackage{booktabs}
\usepackage{adjustbox}
\usepackage{caption}
\usepackage{multirow}

\begin{document}

\maketitle

\begin{abstract}
    Using speaker embeddings as conditioning can strengthen speech enhancement, but most methods either require clean enrollment audio or rely on embeddings extracted from noisy speech, which are fragile under noise and domain shift. We propose G-MaP-SE, a guided enhancement framework that builds a clean-speech embedding prior with a Gaussian Mixture Model (GMM) and refines a noisy conditioning embedding by matching it to this prior. The matched prior embedding is then injected into a time-frequency enhancement backbone via a lightweight gated fusion module. Experiments on VoiceBank+DEMAND and DNS Challenge 2020 datasets show that the proposed prior matching consistently outperforms noisy conditioning and substantially narrows the gap to an oracle clean-conditioning upper bound, while requiring no enrollment audio at inference time. The code, audio samples, and checkpoint are available\footnote{\url{https://github.com/Hello3orld/G-MaP-SE}}.
\end{abstract}

\section{Introduction}
Speech enhancement (SE) aims to improve the perceptual quality and intelligibility of speech signals recorded in everyday acoustic environments, where additive noise and other distortions are inevitable~\cite{Loizou2007SpeechET}. With the progress of deep learning, modern SE systems in both the time domain~\cite{kimSEconformerTimedomainSpeech2021, kongSpeechDenoisingWaveform2022, pascualSEGANSpeechEnhancement2017, pandeyTCNNTemporalConvolutional2019, defossezRealTimeSpeech2020} and the time--frequency (TF) domain~\cite{huDCCRNDeepComplex2020, DBLP:conf/icassp/ZhaoMWG22, caoCMGANConformerbasedMetric2022, luExplicitEstimationMagnitude2025, wangZipEnhancerDualPathDownUp2025} have achieved impressive results on standard benchmarks. Nevertheless, achieving robust enhancement under distribution shift remains challenging, as real-world conditions may differ substantially from the training set in terms of noise types, speaker characteristics, and recording devices~\cite{reddyINTERSPEECH2020Deep2020, saijoInterspeech2025URGENT2025}.

Many recent advances mainly focus on strengthening the SE backbone with more expressive architectures and better training objectives, especially in TF-domain systems that explicitly model magnitude and phase~\cite{luExplicitEstimationMagnitude2025, wangZipEnhancerDualPathDownUp2025}. Other than backbone scaling, robustness is often improved by training on larger and more diverse corpora, applying stronger data augmentation, and designing objectives that better correlate with perceptual quality~\cite{saijoInterspeech2025URGENT2025, DBLP:conf/icml/FuLTL19, fuMetricGANImprovedVersion2021}. While such strategies can improve average performance, they usually require substantial retraining effort and may still degrade when faced with rare or unanticipated conditions~\cite{rehrSNRBasedFeaturesDiverse2021, gonzalezAssessingGeneralizationGap2023}.

Another complementary line of research aims to leverage auxiliary information to better constrain the enhancement process. For example, some methods incorporate visual cues or leverage multi-microphone signals~\cite{blancoAVSEChallengeAudioVisual2023, huangAdvancesMicrophoneArray2025}. In the single-channel setting, a representative formulation is personalized speech enhancement (PSE), where the model is guided by a speaker representation, typically extracted from an enrollment utterance, to better preserve the target speaker and suppress interference~\cite{eskimezPersonalizedSpeechEnhancement2022, juTEAPSETencentEtherealAudioLabPersonalized2022}. This is particularly useful in challenging scenarios such as competing speakers or strong background noise, where purely acoustic cues may be insufficient to maintain speaker consistency. However, clean enrollment audio is often unavailable in practical applications, and requiring users to provide additional recordings also complicates deployment. A more lightweight alternative is to extract the conditioning feature directly from the noisy input~\cite{songExploringWavLMSpeech2023}, but the resulting embedding can be distorted by noise and become unreliable under domain shift, potentially harming the enhancement if used without refinement.

To address this issue, we propose \textbf{G-MaP-SE}, which refines noisy conditioning embeddings via \textbf{G}MM-based \textbf{Ma}tched \textbf{P}rior for guided \textbf{S}peech \textbf{E}nhancement. Specifically, we fit a GMM~\cite{jagtapSpeakerVerificationUsing2015} to clean-speech embeddings offline and, given a noisy utterance, match its embedding to the clean GMM to obtain a refined prior embedding. Intuitively, the clean embedding distribution provides a set of prototypical speaker representations, allowing the noisy embedding to be projected toward a cleaner and more stable region in the embedding space. The key idea is to use the learned clean embedding distribution as a regularizer, yielding a conditioning feature that is more robust to noise corruption. The proposed prior matching module is lightweight and can be integrated into existing speech enhancement backbones through a simple fusion block.

Experiments on the VoiceBank+DEMAND~\cite{valentini-botinhaoInvestigatingRNNbasedSpeech2016} test set and cross-domain evaluation on the DNS Challenge 2020~\cite{reddyINTERSPEECH2020Deep2020} test set demonstrate that the proposed prior matching improves the reliability of noisy conditioning and yields consistent gains under domain shift, without requiring any extra enrollment audio at inference time.

In summary, this work introduces a simple GMM prior matching mechanism for refining noisy conditioning embeddings and validates its effectiveness on both in-domain and cross-domain benchmarks. The proposed module is lightweight and can be used as an add-on component to facilitate guided speech enhancement without modifying the underlying backbone architecture.

\section{Proposed Method}

\subsection{Overall Framework}
\begin{figure}[t]
    \centering
    \includegraphics[width=\linewidth, trim={15pt 0pt 15pt 0pt}, clip]{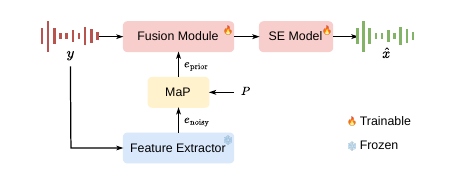}
    \vspace{-10pt}
    \caption{Overview of G-MaP-SE. The noisy input $y$ is fed to both the SE model and a frozen feature extractor. The MaP module matches the noisy embedding $e_{\mathrm{noisy}}$ to a precomputed GMM prior representation $P$ and produces a matched prior embedding $e_{\mathrm{prior}}$. For simplicity, the fusion block is depicted as taking $y$ as input; in practice, fusion is performed on an intermediate SE feature map derived from $y$.}
    \label{fig:gmap_overview}
\end{figure}

Speech enhancement aims at estimating clean speech from noisy observations. Let $T$ denote the number of waveform samples. We use $x\in\mathbb{R}^{T}$ to denote the clean waveform, $y\in\mathbb{R}^{T}$ the noisy waveform, and $n$ additive noise such that $y=x+n$. Given $y$, an SE system outputs an estimate $\hat{x}$.

Figure~\ref{fig:gmap_overview} summarizes G-MaP-SE. The noisy waveform $y$ is fed to a frozen feature extractor to obtain a noisy embedding $e_{\mathrm{noisy}}$. A matching module (MaP) refines $e_{\mathrm{noisy}}$ by matching it to a precomputed GMM prior representation $P$ and outputs a refined prior embedding $e_{\mathrm{prior}}$. The enhancement backbone processes $y$ into intermediate features, and a lightweight fusion block injects $e_{\mathrm{prior}}$ into these intermediate features. The backbone then outputs the enhanced signal $\hat{x}$. This design avoids requiring any additional user-provided enrollment audio at inference time, while providing more reliable conditioning than directly using noisy embeddings. 
In addition, the GMM prior can be swapped across datasets without retraining the enhancement backbone, making it convenient to adapt the conditioning signal to a target domain by simply refitting the prior on available clean speech.

\subsection{GMM Prior Construction}
We construct the GMM prior representation $P$ from embeddings extracted on clean speech waveforms. Specifically, we apply the same feature extractor $f(\cdot)$ as in Figure~\ref{fig:gmap_overview} to a collection of clean utterances and obtain a set of $D$-dimensional embeddings $\{e_i\}_{i=1}^{N}$, where $N$ is the number of clean utterances:
\begin{equation}
    e_i = f(x_i) \in \mathbb{R}^{D}.
\end{equation}
Before fitting the GMM, we apply $\ell_2$ normalization to project embeddings onto the unit hypersphere. This reduces sensitivity to embedding scale and makes the matching geometry at inference time consistent with the prior space:
\begin{equation}
    \tilde{e}_i = \frac{e_i}{\lVert e_i \rVert_2}.
\end{equation}
We then fit a $K$-component Gaussian mixture model to $\{\tilde{e}_i\}$ by maximum likelihood using the expectation--maximization (EM) algorithm~\cite{dempsterMaximumLikelihoodIncomplete1977} as implemented in \texttt{sklearn.mixture.GaussianMixture}\footnote{\url{https://scikit-learn.org/stable/modules/generated/sklearn.mixture.GaussianMixture.html}}:
\begin{equation}
    p(e)=\sum_{k=1}^{K}\pi_k\,\mathcal{N}(e;\mu_k,\Sigma_k),
\end{equation}
where $p(e)$ denotes the probability density of an embedding vector $e\in\mathbb{R}^{D}$ under the mixture model, $\pi_k$ are mixture weights, and $(\mu_k,\Sigma_k)$ are the mean and covariance of each component. We use diagonal covariances for efficiency.

\subsection{MaP}
Given the noisy waveform $y$, we compute an embedding
\begin{equation}
    e_{\mathrm{noisy}} = f(y) \in \mathbb{R}^{D}.
\end{equation}
MaP matches $e_{\mathrm{noisy}}$ to the clean prior $P$. In our implementation, $P$ is represented by the $K$ GMM means $\{\mu_k\}_{k=1}^{K}$. The assignment concentration is controlled by a temperature parameter $\tau$, where smaller values approach hard assignment, and larger values encourage averaging across multiple components. We first normalize embeddings to align the geometry used for GMM fitting:
\begin{equation}
    \tilde{e}=\frac{e_{\mathrm{noisy}}}{\norm{e_{\mathrm{noisy}}}_2}.
\end{equation}
\begin{equation}
    \tilde{\mu}_k=\frac{\mu_k}{\norm{\mu_k}_2}.
\end{equation}
We then compute cosine similarity scores and obtain soft matching weights via a temperature $\tau$, where $\top$ denotes vector transpose:
\begin{equation}
    a_k = \frac{\tilde{e}^\top \tilde{\mu}_k}{\tau}.
\end{equation}
\begin{equation}
    \gamma_k = \frac{\exp(a_k)}{\sum_{j=1}^{K}\exp(a_j)}.
\end{equation}
Finally, the matched prior embedding is obtained by a weighted combination of GMM means:
\begin{equation}
    e_{\mathrm{prior}} = \sum_{k=1}^{K}\gamma_k \mu_k.
\end{equation}

\subsection{Feature Extractor}
\label{sec:feature_extractor}
We use a pretrained embedding model as the feature extractor $f(\cdot)$ and keep it frozen during training. In our implementation, $f(\cdot)$ is an ECAPA-TDNN speaker embedding extractor\footnote{\url{https://wenet.org.cn/downloads?models=wespeaker&version=voxceleb_ECAPA512.onnx}} that outputs $D{=}192$-dimensional embeddings~\cite{desplanquesECAPATDNNEmphasizedChannel2020}. Following the prior construction, we apply the same $\ell_2$ normalization to embeddings. Freezing the extractor keeps the embedding space consistent with the GMM prior and avoids degenerate solutions where the embedding space drifts to overfit the enhancement objective.

\subsection{Fusion Module}
The fusion module injects the matched prior embedding $e_{\mathrm{prior}}$ into the enhancement network. We project the SE feature map and the conditioning embedding with two separate Linear--ReLU projection blocks, and then project both to the same channel dimension. The projected embedding is broadcast along the time and frequency dimensions to match the shape of the SE feature map.

We then compute an element-wise gate from the concatenation of the two projected representations and blend them as follows:
\begin{align}
    g &= \sigma\left(W\,[Y,\,E]\right), \\
    \hat{Y} &= (1-g)\odot Y + g\odot E.
\end{align}
Here, $Y$ and $E$ denote the projected SE feature map and the projected conditioning embedding after broadcast to time--frequency dimensions, respectively. $W$ denotes a learnable linear projection that maps the concatenated features to gate values, $\sigma(\cdot)$ is the sigmoid function, and $\odot$ denotes element-wise multiplication.

\section{Experiments}

\subsection{Dataset}
We conducted experiments on two widely used open-source datasets: VoiceBank+DEMAND (VBD)~\cite{valentini-botinhaoInvestigatingRNNbasedSpeech2016} and the DNS Challenge 2020 dataset (DNS2020)~\cite{reddyINTERSPEECH2020Deep2020}. We train all speech enhancement models on the VBD training split and report results on the VBD test split for in-domain evaluation. To assess cross-domain generalization, we further evaluate the VBD-trained models on the DNS2020 evaluation set without reverberation (DNS2020 w/o reverb).

VBD provides paired clean/noisy utterances built from the VoiceBank corpus~\cite{veauxVoiceBankCorpus2013} and the DEMAND noise database~\cite{thiemannDiverseEnvironmentsMultichannel2013}. The training set contains clean speech from 28 speakers, while the test set contains 2 unseen speakers. Noisy training utterances are generated by mixing clean speech with a set of diverse noise conditions at multiple SNRs, and the test set uses unseen noise types at different SNRs. Following common practice, we resample all audio to 16~kHz in our experiments.

DNS2020 is a large-scale corpus that provides clean speech and noise recordings, along with standardized non-blind evaluation sets consisting of noisy/clean pairs. We use the official evaluation set without reverberation to focus on denoising under domain shift.

To build the clean embedding prior, we extract embeddings from clean utterances and fit a $K$-component GMM in the embedding space. Unless otherwise stated, the prior is learned from the clean utterances in the VBD training split. For cross-domain analysis, we additionally build an alternative prior from DNS2020 clean utterances to better align the prior distribution with the DNS evaluation domain.

\subsection{Experimental Setup}
We follow the official MP-SENet implementation and keep the enhancement backbone architecture unchanged for all systems compared. Specifically, we set the number of channels to 64 and use 4 TF blocks with 4 attention heads. For variants based on conditioning, we use the same frozen ECAPA-TDNN extractor as described in Section~\ref{sec:feature_extractor}. During training, we use oracle conditioning embeddings extracted from clean target speech for all conditioning-based systems to provide a stable conditioning signal and prevent training instability caused by corrupted noisy embeddings. Unless otherwise stated, we set the matching temperature to $\tau{=}0.2$ and use $K{=}192$ mixture components, and the fusion block is inserted after the MP-SENet encoder and before the subsequent sequence modeling blocks.

All audio samples are randomly sliced into 2-second segments. To extract input features from raw waveforms using the short-time Fourier transform (STFT), the FFT size, Hanning window size, and hop size are set to 400, 400, and 100, which correspond to a 25~ms window and a 6.25~ms hop at 16~kHz; consequently, the number of frequency bins is $F{=}201$. The magnitude spectrum compression factor is set to 0.3.

We adopt the same generator loss as MP-SENet~\cite{luExplicitEstimationMagnitude2025}, which is a linear combination of multiple loss terms, including PESQ-based GAN discriminator loss $L_{\mathrm{pesq}}$, STFT consistency loss $L_{\mathrm{stft}}$, magnitude loss $L_{\mathrm{mag}}$, complex-spectrum loss $L_{\mathrm{com}}$, phase loss $L_{\mathrm{pha}}$, and time-domain loss $L_{\mathrm{time}}$:
\begin{equation}
L=\lambda_1 L_{\mathrm{pesq}}+\lambda_2 L_{\mathrm{stft}}+\lambda_3 L_{\mathrm{mag}}+\lambda_4 L_{\mathrm{com}}+\lambda_5 L_{\mathrm{pha}}+\lambda_6 L_{\mathrm{time}}.
\end{equation}
We set $\lambda_1,\lambda_2,\lambda_3,\lambda_4,\lambda_5,\lambda_6$ to 0.05, 0.1, 0.9, 0.1, 0.3, and 0.2, respectively.

The final model has 2.288M trainable parameters, excluding the frozen feature extractor, compared to 2.263M for the original MP-SENet, resulting in an increase of 0.025M parameters introduced by the fusion block. The MaP module itself has no trainable parameters and only performs lightweight matching computations; the GMM prior can be stored as a fixed $K{\times}D$ prototype representation.

We use the AdamW optimizer~\cite{DBLP:conf/iclr/LoshchilovH19} with $\beta_1{=}0.8$, $\beta_2{=}0.99$, and weight decay of 0.01. The learning rate is initialized to 0.0005 and decayed by a factor of 0.99 every epoch. All models are trained on the VBD training split for 500k steps with batch size 4 on a single 32~GB NVIDIA V100 GPU.

\subsection{Evaluation Metrics}
We adopt commonly used objective metrics to assess both speech quality and intelligibility. On VBD, we report wide-band PESQ (WB-PESQ)~\cite{rixPerceptualEvaluationSpeech2001} for perceptual quality, STOI~\cite{taalAlgorithmIntelligibilityPrediction2011} for intelligibility, segmental SNR (SSNR) for noise reduction, and three MOS-predictive composite measures (CSIG, CBAK, and COVL)~\cite{DBLP:journals/taslp/HuL08} that reflect signal distortion, background noise intrusiveness, and overall quality, respectively. On DNS2020, we report WB-PESQ and narrow-band PESQ (NB-PESQ) to evaluate perceptual quality in wide-band and narrow-band settings, STOI to measure intelligibility, and scale-invariant SDR (SI-SDR)~\cite{rouxSDRHalfbakedWell2019} to quantify the distortion between enhanced and clean speech~. For all metrics, higher values indicate better performance.

\begin{table*}[t]
    \captionsetup[table]{aboveskip=0pt, belowskip=8pt}
    \centering
\caption{Results on the VBD test set (in-domain) and the DNS2020 test set without reverberation (cross-domain). All systems are trained on the VBD training set. $^*$ denotes our reproduced result. Oracle-Cond and Noisy-Cond condition the model on embeddings extracted from clean and noisy speech, respectively. G-MaP matches a noisy embedding to the clean GMM prior $P$ (learned from the clean training split of either VBD or DNS2020). Best results are in bold, and second-best are underlined.}
    \label{tab:main_results}
    \begin{adjustbox}{max width=\linewidth}
    \small
    \setlength{\tabcolsep}{4.5pt}
    \begin{tabular}{l c c c c c c c c c c}
        \toprule[1pt]
        \multicolumn{1}{c}{\multirow{2}{*}{\textbf{Model}}} & 
        \multicolumn{6}{c}{\textbf{VBD test set}} &
        \multicolumn{4}{c}{\textbf{DNS2020 w/o reverb test set}} \\
        \cmidrule(lr){2-7}\cmidrule(lr){8-11}
        & \textbf{WB-PESQ} & \textbf{CSIG} & \textbf{CBAK} & \textbf{COVL} & \textbf{STOI (\%)} & \textbf{SSNR (dB)}
        & \textbf{WB-PESQ} & \textbf{NB-PESQ} & \textbf{STOI (\%)} & \textbf{SI-SDR (dB)} \\
        \midrule
        noisy & 1.97 & 3.49 & 2.55 & 2.74 & 92.11 & 1.68 & 1.582 & 2.161 & 91.519 & 9.230 \\
        \midrule
        MP-SENet~\cite{luExplicitEstimationMagnitude2025} & \textbf{3.60} & \textbf{4.81} & \underline{3.99} & \textbf{4.34} & \textbf{96.12} & 10.39 & 2.790 & 3.303 & 95.878 & 16.277 \\
        MP-SENet$^*$ & \underline{3.59} & \underline{4.80} & \textbf{4.00} & \textbf{4.34} & \underline{96.11} & 10.39 & 2.789 & 3.302 & 95.876 & 16.280 \\
        \midrule
        MP-SENet + Oracle-Cond & 3.58 & \underline{4.80} & \textbf{4.00} & \underline{4.33} & 96.05 & \textbf{10.73} & \textbf{2.796} & \textbf{3.352} & \textbf{96.090} & \textbf{16.455} \\
        MP-SENet + Noisy-Cond & 3.56 & 4.79 & \textbf{4.00} & 4.31 & 96.09 & 10.66 & 2.765 & 3.323 & 95.908 & 16.340 \\
        MP-SENet + G-MaP ($P_{\mathrm{VBD}}$) & \underline{3.59} & \underline{4.80} & \textbf{4.00} & \underline{4.33} & 96.10 & \underline{10.67} & \underline{2.794} & 3.349 & 96.065 & \underline{16.454} \\
        MP-SENet + G-MaP ($P_{\mathrm{DNS}}$) & 3.58 & \underline{4.80} & \underline{3.99} & 4.32 & 96.07 & \underline{10.67} & \underline{2.794} & \underline{3.350} & \underline{96.072} & \underline{16.454} \\
        \bottomrule[1pt]
    \end{tabular}
    \end{adjustbox}
    \vspace{-10pt}
\end{table*}

\subsection{Experimental Results}

\subsubsection{Results on VBD and DNS2020}
Table~\ref{tab:main_results} reports the results on the VBD test set (in-domain) and the DNS2020 test set without reverberation (cross-domain), where all systems are trained on the VBD training set. On the VBD test set, the proposed method achieves a performance close to conditioning on embeddings extracted from the clean target speech (oracle conditioning) and improves over conditioning on embeddings extracted from the noisy input (noisy conditioning), while the overall differences remain small. A likely factor is the limited scale of VBD, which constrains the amount and diversity of clean embeddings available for learning a robust prior in the embedding space and therefore limits the impact of prior matching in the in-domain setting.

On the DNS2020 test set, prior matching provides consistent gains across all reported metrics and substantially narrows the gap to oracle conditioning, without requiring any clean enrollment audio at inference time. Moreover, simply replacing the prior learned from the VBD training set with a prior learned from the DNS2020 clean training data further improves performance on the DNS2020 test set without retraining the enhancement backbone, highlighting the plug-and-play nature of the proposed method and the practical benefit of selecting a prior aligned with the target domain.

\begin{figure}[t]
    \centering
    \includegraphics[width=\linewidth]{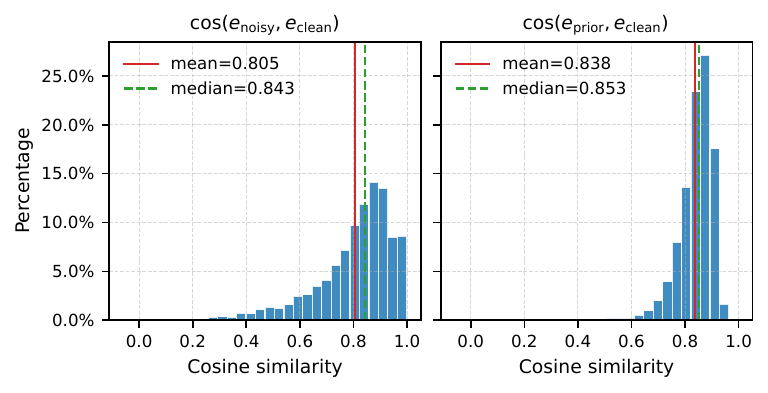}
    \vspace{-10pt}
    \setlength{\belowcaptionskip}{-10pt}
    \caption{Embedding cosine similarity distributions on VBD. Left: $\cos(e_{\mathrm{noisy}}, e_{\mathrm{clean}})$, where $e_{\mathrm{noisy}}$ and $e_{\mathrm{clean}}$ are extracted from the noisy and clean waveforms, respectively. Right: $\cos(e_{\mathrm{prior}}, e_{\mathrm{clean}})$, where $e_{\mathrm{prior}}$ is produced by matching $e_{\mathrm{noisy}}$ to the clean GMM prior. The y-axis denotes the percentage of utterances in each bin.}
    \label{fig:emb_cos_hist}
\end{figure}

\subsubsection{Embedding Refinement Analysis}
Figure~\ref{fig:emb_cos_hist} analyzes the embedding refinement behavior of G-MaP on VBD by comparing the cosine similarity between noisy and clean embeddings and the similarity between the matched prior embedding and the clean embedding. Compared with directly using $e_{\mathrm{noisy}}$, the matched embedding $e_{\mathrm{prior}}$ yields a distribution that shifts toward higher similarity, indicating that GMM matching can correct noise-induced distortions and pull corrupted embeddings closer to the clean embedding space. 

For a subset of utterances, the similarity does not increase after matching, which reflects a limitation of the proposed matching process: the noisy embedding may be assigned to a suboptimal prototype, and therefore does not fully recover the underlying clean embedding. Nevertheless, the matched embedding is computed as a mixture-weighted combination of clean prototypes and thus stays in the clean embedding space, which can preserve useful speaker-related cues while removing noise-induced artifacts; consequently, it is expected to be more reliable than using the noisy embedding alone, especially under noise and domain shift.

\begin{figure}[t]
    \centering
    \includegraphics[width=\linewidth]{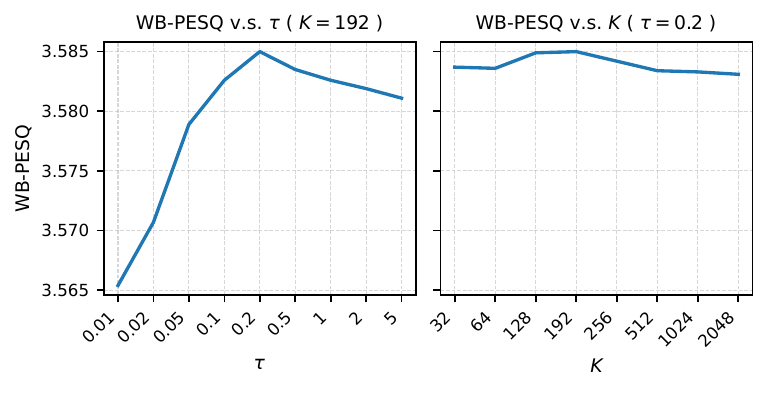}
    \vspace{-10pt}
    \setlength{\belowcaptionskip}{-10pt}
    \caption{Ablation on VBD with respect to the matching temperature $\tau$ and the number of GMM components $K$. Left: in-domain WB-PESQ versus $\tau$ with $K{=}192$. Right: in-domain WB-PESQ versus $K$ with $\tau{=}0.2$.}
    \label{fig:tau_k_ablation}
\end{figure}

\subsubsection{Ablation Study}
Figure~\ref{fig:tau_k_ablation} shows the ablation results on VBD with respect to the matching temperature $\tau$ and the number of GMM components $K$. As $\tau$ increases from very small values, performance first improves, peaks around $\tau{=}0.2$, and then slightly degrades for larger $\tau$. This trend is consistent with the role of $\tau$ in soft matching: overly small $\tau$ makes the assignment close to hard selection and may amplify embedding noise by relying on a single prototype, whereas overly large $\tau$ over-smooths the weights and approaches an averaged prototype that is less discriminative. The best performance is achieved by balancing robustness and specificity.

When varying $K$ with $\tau{=}0.2$, performance exhibits a mild peak around $K{=}192$. Increasing $K$ from small values improves the granularity of the clean prior and provides a richer set of matching prototypes. However, with limited data for fitting the prior, excessively large $K$ can lead to poorly estimated components and reduced effective coverage, slightly affecting performance. Compared with $K$, performance is more sensitive to $\tau$, suggesting that assignment softness plays a larger role than the exact number of prototypes within a reasonable range.

\section{Conclusion}
We proposed G-MaP-SE, a guided speech enhancement framework that refines noisy conditioning embeddings by matching them to a GMM prior learned from clean speech and injects the matched prior embedding into a TF-domain enhancement backbone via gated fusion. Experiments on VoiceBank+DEMAND and DNS Challenge 2020 datasets show that prior matching improves robustness under noise and domain shift, narrowing the gap to oracle clean conditioning without requiring enrollment audio at inference time. Future work will explore building stronger and more domain-adaptive priors using larger and more diverse clean-speech corpora, developing more accurate matching strategies that better preserve speaker characteristics under severe noise, and designing more effective fusion mechanisms to further improve robustness and generalization.

\section{Generative AI Use Disclosure}
All (co-)authors are responsible and accountable for the work and the content of this paper, and they consent to its submission. No generative AI tool is listed as a co-author. Generative AI tools were used only for editing and polishing the manuscript and were not used to produce any significant part of the manuscript or generate the core scientific content, including the proposed method, experiments, results, or conclusions.

\bibliographystyle{IEEEtran}
\bibliography{mybib}

\end{document}